\documentclass[aps,prd,amsmath,
twocolumn,
showpacs,preprintnumbers,
superscriptaddress,nofootinbib]{revtex4}

\usepackage{amsmath}
\usepackage{graphicx}
\usepackage{amssymb}
\usepackage{amsthm, amscd}
\usepackage{amsfonts}
\usepackage{subfigure}
\usepackage{hyperref}
\usepackage[usenames,dvipsnames]{color}
\usepackage[squaren]{SIunits}
\usepackage{ulem} 

\newcommand{\bw}{\begin{widetext}}
\newcommand{\ew}{\end{widetext}}

\newcommand{\ee}{\mathrm{e}}
\newcommand{\aye}{\mathrm{i}}
\newcommand{\Msun}{M_{\odot}}

\addunit{\mebi}{Mi}
\addunit{\byte}{B}

\newcommand{\CITA}{Canadian Institute for Theoretical Astrophysics, 60 St.\ George Street, University of Toronto, Toronto, ON M5S 3H8, Canada}
\newcommand{\Perimeter}{Perimeter Institute for Theoretical Physics, Waterloo, Ontario N2L 2Y5, Canada}
\newcommand{\AEI}{Albert-Einstein-Institut, Max-Planck-Institut f\"ur Gravitationsphysik, D-30167 Hannover, Germany}
\newcommand{\Leibniz}{Leibniz Universit\"at Hannover, D-30167 Hannover, Germany}
\newcommand{\UofTAstro}{Department of Astronomy and Astrophysics, 50 St.\ George Street, University of Toronto, Toronto, ON M5S 3H4, Canada}

\begin{document}

\title{Interpolation in waveform space:  enhancing the accuracy of gravitational waveform families using numerical relativity}

\author{Kipp Cannon}
\email{kipp.cannon@ligo.org}
\affiliation{\CITA}
\author{J.D.\ Emberson}
\email{emberson@astro.utoronto.ca}
\affiliation{\CITA}
\affiliation{\UofTAstro}
\author{Chad Hanna}
\email{chad.hanna@ligo.org}
\affiliation{\Perimeter}
\author{Drew Keppel}
\email{drew.keppel@ligo.org}
\affiliation{\AEI}
\affiliation{\Leibniz}
\author{Harald P.\ Pfeiffer}
\email{pfeiffer@cita.utoronto.ca}
\affiliation{\CITA}


\begin{abstract}
Matched-filtering for the identification of compact object mergers in
gravitational-wave antenna data involves the comparison of the data stream
to a bank of template gravitational waveforms.  Typically the template bank
is constructed from phenomenological waveform models since these can be
evaluated for an arbitrary choice of physical parameters. Recently it has
been proposed that singular value decomposition (SVD) can be used to reduce
the number of templates required for detection.  As we show here, another
benefit of SVD is its removal of biases from the phenomenological templates
along with a corresponding improvement in their ability to represent
waveform signals obtained from numerical relativity (NR) simulations. Using
these ideas, we present a method that calibrates a reduced SVD basis of
phenomenological waveforms against NR waveforms in order to construct a new
waveform approximant with improved accuracy and faithfulness compared to
the original phenomenological model. The new waveform family is given
numerically through the interpolation of the projection coefficients of NR
waveforms expanded onto the reduced basis and provides a generalized scheme
for enhancing phenomenological models.
\end{abstract}

\preprint{LIGO-P1200138}

\maketitle

\section{Introduction}

Developments are currently underway to promote the sensitivity of LIGO and
to improve its prospect for detecting gravitational waves emitted by
compact object binaries \cite{abbottetal/ligo:2009,adlreferencedesign}.  Of
particular interest are the detection of gravitational waves released
during the inspiral and merger of binary black hole (BBH) systems.
Detection rates for BBH events are expected to be within 0.4--1000 per year
with Advanced LIGO \citep{abadie/etal:2010}. It is important that rigorous detection 
algorithms be in place in order to maximize the number of detections of gravitational
wave signals. 

The detection pipeline currently employed by LIGO involves a matched-filtering
process whereby signals are compared to a pre-constructed template bank of
gravitational waveforms. The templates are chosen to cover some interesting
region of mass-spin parameter space and are placed throughout it in such a
way that guarantees some minimal match between any arbitrary point in
parameter space and its closest neighbouring template. Unfortunately, the
template placement strategy generally requires many thousands of templates
(e.g.\ \citep{abadie/etal:2011}) evaluated at arbitrary mass and spin;
something that cannot be achieved using the current set of numerical
relativity (NR) waveforms.

To circumvent this issue, LIGO exploits the use of analytical waveform
families like phenomenological models
\cite{Ajith:2009bn,santamaria/etal:2010} or effective-one-body models
\cite{Pan:2011gk,Taracchini:2012ig}.  We shall focus here on the
Phenomenological B (PhenomB) waveforms developed by
\cite{santamaria/etal:2010}.  This waveform family describes BBH systems
with varying masses and aligned-spin magnitudes (i.e.\ non-precessing
binaries).  The family was constructed by fitting a parameterized model to
existing NR waveforms in order to generate a full inspiral-merger-ringdown
(IMR) description as a function of mass and spin.  The obvious appeal of
the PhenomB family is that it allows for the inexpensive construction of
gravitational waveforms at arbitrary points in parameter space and can thus
be used to create arbitrarily dense template banks. 

To optimize computational efficiency of the detection process it is desirable to reduce
the number of templates under consideration. A variety of reduced bases
techniques have been developed, either through singular-value decomposition
(SVD) \cite{cannon/etal:2010,cannon/etal:2011b}, or via a greedy
algorithm \cite{Field:2011mf}. SVD is an algebraic manipulation that
transforms template waveforms into an orthonormal basis with a prescription
that simultaneously filters out any redundancies existing within the
original bank.  As a result,  the number of templates required for
matched-filtering can be significantly reduced.  In addition, it has been
shown in \cite{cannon/etal:2011c} that, upon projecting template waveforms
onto the orthonormal basis produced by the SVD, interpolating the
projection coefficients provides accurate approximations of other IMR
waveforms not included in the original template bank.

In this paper, we continue to explore the use of the interpolation of
projection coefficients.  We take a novel approach that utilizes both the
analytic PhenomB waveform family \cite{santamaria/etal:2010} and NR hybrid
waveforms \cite{Scheel2009,Buchman:2012dw,MacDonald:2011ne}.  We apply SVD to a
template bank constructed from an analytical waveform family to construct an
orthonormal basis spanning the waveforms, then project the NR waveforms onto
this basis and interpolate the projection coefficients to allow arbitrary
waveforms to be constructed, thereby obtaining a new waveform approximant.  We
show here that this approach improves upon the accuracy of the original
analytical waveform family.  The original waveform family shows mismatches with
the NR waveforms as high as $0.1$ when no extremization over physical
parameters is applied (i.e., a measure of the ``faithfulness" of the
waveform approximant), and mismatches of $0.02$ when maximized over total mass
(i.e., a measure of the ``effectualness" of the waveform approximant).
With our SVD accuracy booster, we are able to construct a new waveform family
(given numerically) with mismatches $<0.005$ even without extremization over
physical parameters.

This paper is organized as follows. We begin in Section \ref{sec:mbias} where we 
provide definitions to important terminology used in our paper. We then compare our
NR hybrid waveforms to the PhenomB family and show that a mass-bias exists between
the two. In Section \ref{sec:method} we present our SVD
accuracy booster applied to the case study of equal-mass, zero-spin binaries.
In Section \ref{sec:2d} we investigate the feasibility of extending this approach to 
include unequal-mass binaries.
We finish with concluding remarks in Section \ref{sec:discussion}. 

\section{Gravitational Waveforms}
\label{sec:mbias}

\subsection{Terminology}
\label{sec:mbiasA}

A gravitational waveform is described through a complex function, ${\bf
h}(t)$, where real and imaginary parts store the sine and cosine components
of the wave. The specific form of ${\bf h}(t)$ depends on the parameters of
the system, in our case the total mass $M = m_1 + m_2$ and the mass-ratio $q=m_1/m_2$.
While ${\bf h}(t)$ is a continuous function of time, we discretize by
sampling ${\bf h}(t_i)$, where the sampling times $t_i$ have uniform
spacing $\Delta t = \unit{2^{-15}}{\second}$.

We shall also whiten any gravitational waveform ${\bf h}(t)$.  
This processes is carried out in frequency space via
\begin{equation}
\tilde{{\bf h}}_{\rm w}(f) = \frac{\tilde{{\bf h}}(f)}{\sqrt{S_n(f)}},
\label{eq:whiten}
\end{equation} 
where $S_n(f)$ is the LIGO noise curve and $\tilde{{\bf h}}(f)$ is the
Fourier transform of ${\bf h}(t)$. The whitened time-domain waveform, ${\bf
h}_{\rm w}(t)$, is obtained by taking the inverse Fourier transform of
\eqref{eq:whiten}.  In the remainder of the paper, we shall always refer to whitened waveforms, dropping the subscript ``w''. For our purposes it suffices to take $S_n(f)$ to be the Initial LIGO noise curve. Using the Advanced LIGO noise curve would only serve to needlessly complicate our approach by making waveforms longer in the low frequency domain.

As a measure of the level of agreement between two waveforms, ${\bf h}(t)$ and ${\bf g}(t)$, we will use their match, or overlap, $\mathcal{O}({\bf h}, {\bf g})$~\cite{cutlerflanagan:1994,Balasubramanian:1996,Owen:1996}.  We define
\begin{equation}
\mathcal{O}({\bf h}, {\bf g}) \equiv 
\max_{\Delta T}\left| \frac{\langle {\bf h}, {\bf g} \rangle}{||{\bf h}|| \cdot  ||{\bf g}||} \right|,
\label{eq:overlap}
\end{equation}
where $\langle {\bf h}, {\bf g} \rangle$ is the standard complex inner product and the norm $||{\bf h}|| \equiv \sqrt{\langle {\bf h}, {\bf h} \rangle}$.  We always consider the overlap maximized over time- and phase-shifts between the two waveforms.  The time-maximization is indicated in \eqref{eq:overlap}, and the phase-maximization is an automatic consequence of the modulus.
Note that $0 \leq \mathcal{O}({\bf h}, {\bf g}) \leq 1$. For discrete sampling at points $t_i = t_0 + i \Delta t$ we have that
\begin{equation}
\langle {\bf h}, {\bf g} \rangle = \sum_i {\bf h}(t_i) \cdot {\bf g}^*(t_i), 
\label{eq:inner}
\end{equation}
where ${\bf g}^*(t)$ is the complex conjugate of ${\bf g}(t)$.  Without
whitening, \eqref{eq:inner} would need to be evaluated in the frequency
domain with a weighting factor $1/S_n(f)$.  The primary advantage of
\eqref{eq:inner} is its compatibility with formal results for the SVD, which will allow us to make more precise statements below.
When maximizing over time-shifts $\Delta T$, we ordinarily 
consider discrete time-shifts in integer multiples of $\Delta t$, as this avoids interpolation.
After the overlap has been maximized, it is useful to speak in terms of the mismatch, $\mathcal{M}({\bf h}, {\bf g})$, 
defined simply as
\begin{equation}
\mathcal{M}({\bf h}, {\bf g}) \equiv 1 - {\cal O}({\bf h}, {\bf g}).
\label{eq:error}
\end{equation}
We use this quantity throughout the paper to measure the level of disagreement between waveforms. 

\subsection{NR Hybrid Waveforms}

We use numerical waveforms computed with the Spectral Einstein Code {\tt
SpEC} \cite{SpECWebsite}.  Primarily, we use the 15-orbit equal-mass (mass-ratio $q=1$),
zero-spin (effective spin $\chi=0$) waveform described in \cite{Boyle2007,Scheel2009}.  In Section
\ref{sec:2d}, we also use unequal mass waveforms computed by
\cite{Buchman:2012dw}.  The waveforms are hybridized with a TaylorT3
post-Newtonian (PN) waveform as described
in \cite{MacDonald:2011ne,MacDonald:2012inPrep} at matching frequencies
$M\omega=0.038,0.038,0.042,0.044,$ and $0.042$ for mass-ratios $q=1,2,3,4,$ and $6$,
respectively. 

TaylorT4 at 3.5PN order is known to match NR simulations
exceedingly well for equal-mass, zero-spin BBH systems~\cite{Boyle2007} (see also Fig.~9 of~\cite{MacDonald:2011ne}).
For $q\!=\!1$, a TaylorT3 hybrid is very similar to a TaylorT4 hybrid, cf.\
Figure 12 of \cite{MacDonald:2011ne}.  The mismatch between TaylorT3 and
TaylorT4 hybrids
is below $10^{-3}$ at $M=\unit{10}{\Msun}$, dropping to below $10^{-4}$ for
$\unit{15}{\Msun}\le M\le \unit{20}{\Msun}$, and 
$10^{-5}$ for $\unit{20}{\Msun}\le M\le \unit{100}{\Msun}$.  
These mismatches are significantly smaller than mismatches arising in the study
presented here, so we conclude that our results are not influenced by the
accuracy of the utilized $q=1$ PN-NR hybrid waveform.  For higher mass-ratios, the PN-NR hybrids have a larger error due to the post-Newtonian waveform~\cite{MacDonald:2012inPrep}.  The error-bound on the hybrids increases with mass-ratio, however, is mitigated in our study here, because  we use the $q\ge 2$ hybrids only for total mass of $\unit{50}{\Msun}$, where less of the post-Newtonian waveform is in band.

Because NR simulations are not available for arbitrary mass ratios, we will
primarily concentrate our investigation to the equal-mass and zero-spin NR
hybrid waveforms described above.  The full IMR waveform can be generated
at any point along the $q = 1$ line through a simple rescaling of amplitude
and phase with total mass $M$ of the system.  Despite such a simple
rescaling, the $q = 1$ line lies orthogonal to lines of constant
chirp mass \citep{allen/etal:2005}, therefore tracing a steep gradient in
terms of waveform overlap, and encompassing a large degree of waveform
structure.

\subsection{PhenomB Waveforms}
\label{sec:mbiasB}

\begin{figure}
\resizebox{\linewidth}{!}{\includegraphics{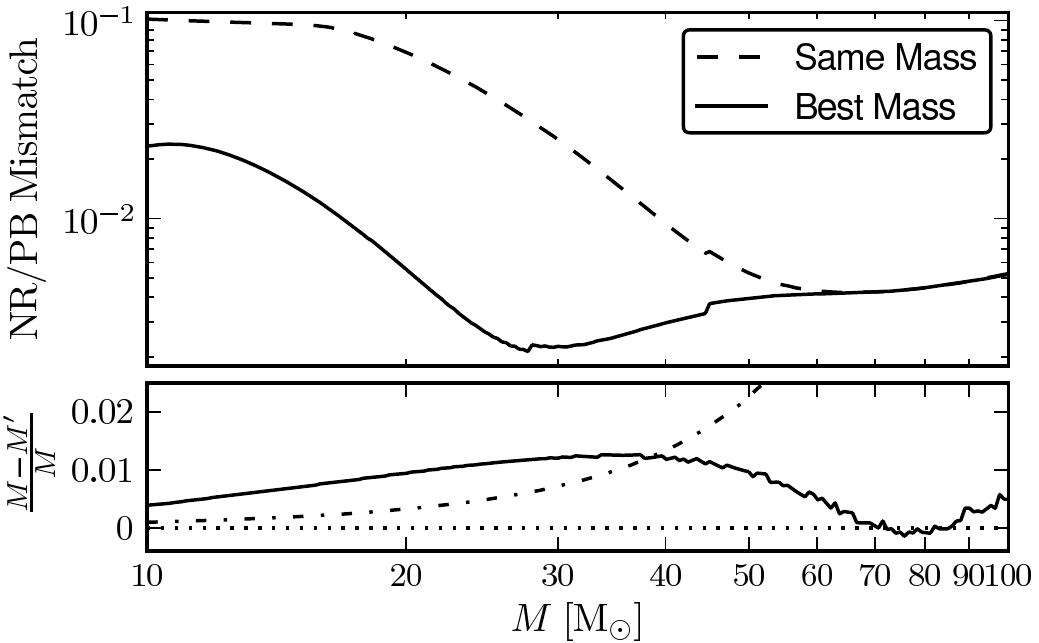}}
\caption{The dashed line in the top panel traces the mismatch between equal-mass, zero-spin NR and PhenomB waveforms of the same total mass $M$. Mismatch is reduced (solid line) by searching to find the mass $M^\prime$ for which $\mathcal{M}[{\bf h}_{\rm NR}(M), {\bf h}_{\rm PB}(M^\prime)]$ is a minimum. We generally find that $M^\prime < M$ as shown in the bottom panel where the solid line traces the mass-bias $(M - M^\prime)/ M$. For comparison, the dot-dashed curve in the bottom panel traces the mass spacing $(M_k - M_{k-1}) / M_k$ for a template bank of PhenomB waveforms satisfying $\mathcal{M}[{\bf h}_{\rm PB}(M_{k-1}), {\bf h}_{\rm PB}(M_k)] = 8^{-2}/2$.}
\label{figure:bias}
\end{figure}

Since our procedure for constructing an orthonormal basis begins with
PhenomB waveforms, let us now investigate how well these waveforms model
the NR waveforms to be interpolated.  For this purpose, we adopt the
notation ${\bf h}_{\rm NR}(M)$ and ${\bf h}_{\rm PB}(M)$ to represent NR
and PhenomB waveforms of total mass $M$, respectively.  We quantify the
faithfulness of the PhenomB family by computing the mismatch
$\mathcal{M}[{\bf h}_{\rm NR}(M), {\bf h}_{\rm PB}(M)]$ as a function of
mass. The result of this calculation for $\unit{10}{\Msun} \leq M \leq
\unit{100}{\Msun}$ is shown as the dashed curve in the top panel of Figure
\ref{figure:bias}. The mismatch starts off rather high with $\mathcal{M}
\approx 0.1$ at $\unit{10}{\Msun}$ and then slowly decreases as the mass is
increased, until eventually flattening to $\mathcal{M} \approx 0.005$ at
high mass.

The mismatch between NR and PhenomB waveforms can be reduced by optimizing
over a mass-bias. This is accomplished by searching for the mass $M^\prime$
for which the mismatch $\mathcal{M}[{\bf h}_{\rm NR}(M), {\bf h}_{\rm
PB}(M^\prime)]$ is a minimum.  The result of this process is shown by the
solid line in the top panel of Figure \ref{figure:bias}.  Allowing for a
mass bias significantly reduces the mismatch for $M\lesssim
\unit{50}{\Msun}$.  The mass $M'$ that minimizes mismatch is generally
smaller than the mass $M$ of our NR ``signal'' waveform, $M^\prime < M$
over almost all of the mass range considered.  Apparently, PhenomB
waveforms are systematically underestimating the mass of the ``true'' NR
waveforms, at least along the portion of parameter space considered here.
The solid line in the bottom panel of Figure \ref{figure:bias} plots the
relative mass-bias, $(M-M')/M$.  At $\unit{10}{\Msun}$ this value is
$0.3\%$, and it rises to just above $1\%$ for
$\unit{30}{\Msun}$.\footnote{In our calculation we have fixed $q=1$ and
$\chi=0$. A more comprehensive minimization over mass, mass ratio, and
effective spin might change this result.}

It is useful to consider how this mass bias compares to the potential
parameter estimation accuracy in an early detection.  For a signal with a
matched-filter signal-to-noise ratio (SNR) of 8 --- characteristic of early
detection scenarios --- template/waveform mismatches will influence
parameter estimation when the mismatch is $\mathcal{M} \geq 8^{-2}/2\sim 0.01$
\cite{Ohme:2012}.  Placing a horizontal cut on the top panel of Figure~\ref{figure:bias} at $\mathcal{M}=8^{-2}/2$, we see that for $M\gtrsim \unit{40}{\Msun}$ PhenomB waveform errors have no observational consequence; for $\unit{15}{\Msun}\lesssim M\lesssim \unit{40}{\Msun}$ a PhenomB waveform with the wrong mass will be the best match for the signal.  For $M\lesssim \unit{15}{\Msun}$ the missmatch between equal-mass PhenomB waveforms and NR (when optimizing over mass) grows to $\sim 0.02$.  Optimization over mass-ratio will reduce this mismatch, but we have not investigated to what degree.

\section{Interpolated Waveform Family}
\label{sec:method}

\subsection{PhenomB Template Bank}

We aim to construct an orthonormal basis via the SVD of a bank of PhenomB
template waveforms, and then interpolate the coefficients of NR waveforms
projected onto this basis to generate a waveform family with improved NR
faithfulness.  The first step is to construct a template bank of PhenomB
waveforms, with attention restricted to equal-mass, zero-spin binaries. An
advantage of focusing on the $q = 1$ line is that template bank
construction can be simplified by systematically arranging templates in
ascending order by total mass. 

With this arrangement we define a template bank to consist of $N$ PhenomB waveforms, labelled
${\bf g}_{i} \equiv {\bf h}_{\rm PB}(M_i)$ ($i = 1, 2, \dots, N$), with $M_{i+1} > M_{i}$ 
and with adjacent templates satisfying the relation:
\begin{equation}
|\mathcal{O}^\prime- \mathcal{O}({\bf g}_i, {\bf g}_{i+1})| \le \varepsilon,
\label{eq:tempover}
\end{equation}
where $\mathcal{O}^\prime$ is the desired overlap between templates and 
$\varepsilon$ is some accepted tolerance in this value. 
The template bank is
initiated by choosing a lower mass bound $M_1 = M_{\rm min}$ and assigning
${\bf g}_1 = {\bf h}_{\rm PB}(M_1)$.
Successive templates are found by sequentially moving toward higher mass in
order to find waveforms satisfying \eqref{eq:tempover} until some maximum mass $M_{\rm max}$ is reached.  
Throughout each trial, overlap between waveforms is maximized continuously over phase shifts and discretely 
over time shifts. For template bank construction we choose to refine the optimization over time by considering 
shifts in integer multiples of $\Delta t / 100$.

We henceforth refer to our fiducial template bank which employs the parameters 
$M_{\rm min} = \unit{15}{\Msun}$, $M_{\rm max} = \unit{100}{\Msun}$, 
$\mathcal{O}^\prime = 0.97$, and $\varepsilon = 10^{-12}$. The lower mass bound 
was chosen in order to obtain a reasonably sized template bank containing 
$N = 127$ waveforms; pushing downward to $\unit{10}{\Msun}$ results in more than
doubling the number of templates.  Template waveforms each have a
duration of \(\unit{8}{\second}\) and are uniformly sampled at $\Delta t =
\unit{2^{-15}}{\second}$ (a sample frequency of $\unit{32768}{\hertz}$).  \(\unit{508}{\mebi\byte}\) of memory is required to store this template bank using double-precision waveforms.


\subsection{Representation of Waveforms in a Reduced SVD Basis}

The next step is to transform the template waveforms into an orthonormal basis. Following the presentation in \cite{cannon/etal:2010}, this is achieved by arranging the templates into the rows of a matrix ${\bf G}$ and factoring through SVD to obtain
\begin{equation}
{\bf G} = {\bf V \Sigma U}^{\rm T},
\label{eq:SVD}
\end{equation}
where ${\bf U}$ and ${\bf V}$ are orthogonal matrices and ${\bf \Sigma}$ is a diagonal matrix whose non-zero elements along the main diagonal are referred to as singular values. The SVD for ${\bf G}$ is uniquely defined as long as the singular values are arranged in descending order along the main diagonal of ${\bf \Sigma}$. 

The end result of \eqref{eq:SVD} is to convert the $N$ complex-valued templates into $2N$ real-valued orthonormal basis waveforms. The $k^{\rm th}$ basis waveform, ${\bf u}_k$, is stored in the $k^{\rm th}$ row of ${\bf U}$, and associated with this mode is the singular value, $\sigma_k$, taken from the $k^{\rm th}$ element along the main diagonal of ${\bf \Sigma}$. 
One of the appeals of SVD is that the singular values rank the basis waveforms with respect to their ability to represent the original templates. This can be exploited in order to construct a reduced basis that spans the space of template waveforms to some tolerated mismatch.

For instance, suppose we choose to reduce the basis by considering only the first $N^\prime < 2N$ basis modes while discarding the rest. Template waveforms can be represented in this reduced basis by expanding them as the sum
\begin{equation}
{\bf g}^\prime = \sum_{k=1}^{N^\prime} \mu_k {\bf u}_k,
\label{eq:gproj}
\end{equation}
where $\mu_k$ are the complex-valued projection coefficients,
\begin{equation}
\mu_k \equiv \langle {\bf g}, {\bf u}_k \rangle.
\label{eq:gprojc}
\end{equation}
The prime in \eqref{eq:gproj} is used to stress that the reduced basis is generally unable to fully represent the original template.\footnote{In the case where $N^\prime = 2N$ we are guaranteed from \eqref{eq:SVD} to completely represent the template.}
It was shown in \citep{cannon/etal:2010} that the mismatch expected from reducing the basis in this way is
\begin{equation}
\langle \mathcal{M} \rangle \equiv
\langle \mathcal{M}({\bf g}^\prime, {\bf g}) \rangle
= \frac{1}{4N} \sum_{k = N^\prime + 1}^{2N} \sigma_k^2.
\label{eq:svderror}
\end{equation}
Given ${\bf \Sigma}$, \eqref{eq:svderror} can be inverted to determine the number of basis waveforms, $N^\prime$,  required to represent the original templates for some expected mismatch $\langle \mathcal{M} \rangle$. 

\eqref{eq:svderror} provides a useful estimate to the mismatch in
represeting templates from a reduced SVD basis. In order to investigate its
accuracy, however, we should compute the mismatch explicitly for each template
waveform. Using the orthonormality condition $\langle {\bf
u}_j, {\bf u}_k \rangle = \delta_{jk}$, it is easy to show from \eqref{eq:gproj} that the mismatch 
between the template and its projection can be expressed in terms of the projection coefficients:
\begin{equation}
\mathcal{M}({\bf g}^\prime, {\bf g}) = 
1 - \sqrt{\sum_{k=1}^{N^\prime} \mu_k \mu_k^*}.
\label{eq:grecerror}
\end{equation}
This quantity is minimized continuously over phase and discretely over time 
shifts in integer multiples of $\Delta t$.

\begin{figure}
\resizebox{\linewidth}{!}{\includegraphics{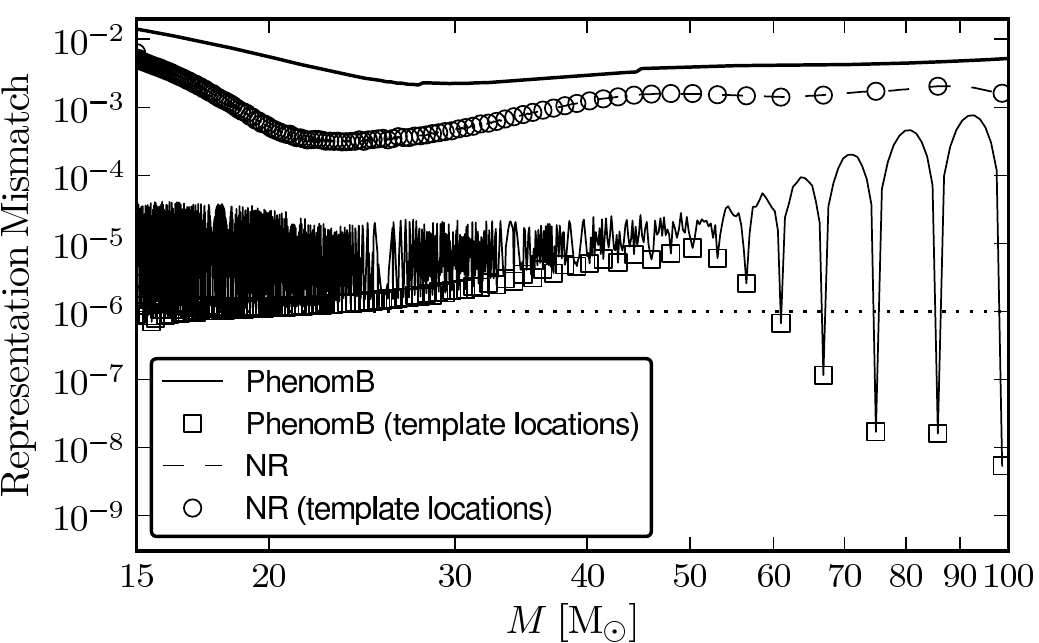}}
\caption{Representation mismatch for the reduced SVD basis with expected tolerance $\langle \mathcal{M} \rangle = 10^{-6}$ (traced by dotted line). Open squares (open circles) show the representation mismatch for PhenomB (NR) waveforms evaluated at the PhenomB template locations, while the thin solid line (thin dashed line) traces representation mismatch of PhenomB (NR) waveforms evaluated between templates. The NR waveforms cannot be represented as well as their PhenomB counterparts, although their total match is improved over using PhenomB waveforms alone. This is evidenced by the thick solid line tracing the mass-optimized mismatch between NR and PhenomB waveforms (i.e.\ the solid line in the top panel of Figure \ref{figure:bias}).}
\label{figure:rec}
\end{figure}

Choosing $\langle\mathcal{M}\rangle = 10^{-6}$,
\eqref{eq:svderror} predicts that $N^\prime = 123$ of the $2N = 254$ basis
waveforms from our fiducial template bank are required to represent the templates
to the desired accuracy. In Figure \ref{figure:rec} we compare the expected
mismatch of $10^{-6}$ to the actual mismatches computed from \eqref{eq:grecerror} 
for each PhenomB waveform in the template bank. The open squares in this plot
show that the actual template mismatch has a significant amount of scatter
about $\langle \mathcal{M} \rangle$, but averaged over a whole remains 
well bounded to the expected result. The PhenomB template waveforms can thus be
represented to a high degree from a reasonably reduced SVD basis.


We are of course more interested in determining how well NR waveforms can be 
represented by the same reduced basis of PhenomB waveforms. 
Since NR and PhenomB waveforms are not equivalent, 
\eqref{eq:svderror} cannot be used to estimate the mismatch obtained when projecting NR waveforms
onto the reduced basis. We must therefore compute their representation mismatch explicitly. A general
waveform, ${\bf h}$, can be represented by the reduced basis in analogy to 
\eqref{eq:gproj} by expressing it as the sum:
\begin{equation}
{\bf h}^\prime = \sum_{k = 1}^{N^\prime} \mu_k {\bf u}_k,
\label{eq:hproj}
\end{equation}
where $\mu_k = \langle {\bf h}, {\bf u}_k \rangle$. As before, the represented waveform ${\bf h}^\prime$
will in general be neither normalized nor equivalent to the original waveform ${\bf h}$. The mismatch
between them is
\begin{equation}
\mathcal{M}({\bf h}^\prime, {\bf h})
= 1 - \sqrt{\sum_{k=1}^{N^\prime} \mu_k \mu_k^*},
\label{eq:recerror}
\end{equation}
where we remind the reader that we always minimize over continuous phase
shifts and discrete time shifts of the two waveforms. 

\begin{figure}
\resizebox{\linewidth}{!}{\includegraphics{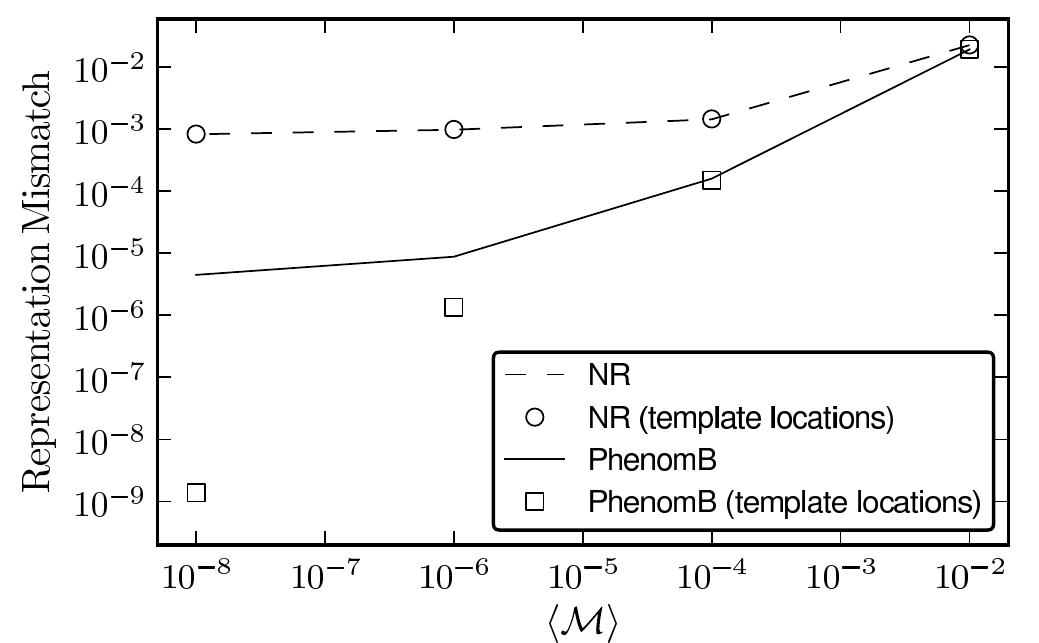}}
\caption{Convergence of representation mismatch with tightening tolerance $\langle {\cal M}\rangle$. As a function of $\langle {\cal M}\rangle$, we plot the averages of the four data-sets shown in Figure \ref{figure:rec}.  The representation mismatch of the PhenomB waveforms for masses {\em in} the template bank used to construct the SVD basis decays roughly with $\langle {\cal M}\rangle$. The representation mismatch of PhenomB waveforms for masses {\em between} the masses in the template bank reaches a plateau of $\sim 10^{-5}$ (the precise value depends on $\mathcal{O}^\prime$, cf. \eqref{eq:tempover}).  The representation mismatch of NR waveforms is yet larger with a plateau of $\sim 10^{-3}$ that flattens for larger $\langle \mathcal{M} \rangle$ (this flattening depends only mildly on $\mathcal{O}^\prime$).}
\label{figure:trun}
\end{figure}

In Figure \ref{figure:rec} we use open circles to plot the representation mismatch of NR waveforms evaluated
at the same set of masses $M_i$ from which the PhenomB template bank was constructed. 
We see that NR waveforms can be represented in the reduced basis with a mismatch less than $10^{-3}$ over
most of the template bank boundary. This is about a factor of five improvement in what can be achieved by
using PhenomB waveforms optimized over mass. Since NR waveforms were not originally included
in the template bank, and because a mass-bias exists between the PhenomB waveforms which were included,
we can expect that the template locations have no special meaning to NR waveforms. This is evident from the 
thin dashed line which traces the NR representation mismatch for masses evaluated between the discrete templates. 
This line varies smoothly across the considered mass range and exhibits no special features at the template locations.
This is in contrast to the thin solid line which traces PhenomB representation mismatch evaluated between templates.
In this case, mismatch rises as we move away from one template and subsequently falls back down as the next template
is approached.

The representation tolerance $\langle{\cal M}\rangle$ of the SVD is a free parameter, which so far, 
we have constrained to be $\langle{\cal M}\rangle = 10^{-6}$. When this tolerance is varied, we observe
the following trends: (i) PhenomB representation mismatch generally follows $\langle\mathcal{M}\rangle$;
(ii) NR representation mismatch follows $\langle\mathcal{M}\rangle$ at first and then {\em saturates}
to a minimum as the representation tolerance is continually reduced. These trends are observed in Figure \ref{figure:trun}
where we plot NR and PhenomB representation mismatch averaged over the mass boundary of the template bank evaluated both
at and between templates. The saturation in NR representation mismatch
occurs when the reduced basis captures all of the NR waveform structure contained within the PhenomB basis. 
Reducing the basis further hits a point of diminishing returns as the increased computational cost associated with
a larger basis outweighs the benefit of marginally improving NR match. 

\subsection{Interpolation of NR Projection Coefficients}

We now wish to examine the possibility of using the reduced SVD basis of PhenomB
template waveforms to construct a new waveform family with improved NR representation. 
The new waveform family would
be given by a numerical interpolation of the projection coefficients of NR waveforms
expanded onto the reduced basis. Here we test this using the fiducial template bank and reduced
basis described above.

The approach is to sample NR projection coefficients, $\mu_k(x_i)$, at some set of locations, $x_i$, 
and then perform an interpolation to obtain the continuous function $\mu_k^\prime(x)$ that 
can be evaluated for arbitrary $x$. The accuracy of the interpolation scheme is maximized by finding 
the space for which $\mu_k(x_i)$ are smooth functions of $x$. It is reasonable to suppose that the projection 
coefficients will vary on a similar scale over which the waveforms themselves vary. Hence, a suitable 
space to sample along is the space of constant waveform overlap. We define this to be the space 
$x = [-1, 1]$ for which the physical template masses are mapped according to:
\begin{equation}
M_i \rightarrow x_i = -1 + 2 \frac{i-1}{N-1}. 
\label{eq:xspace}
\end{equation}
Moving a distance $\Delta x = 2/(N-1)$ in this space is thus equivalent to moving a distance equal 
to the overlap between adjacent templates. 

\begin{figure}
\resizebox{\linewidth}{!}{\includegraphics{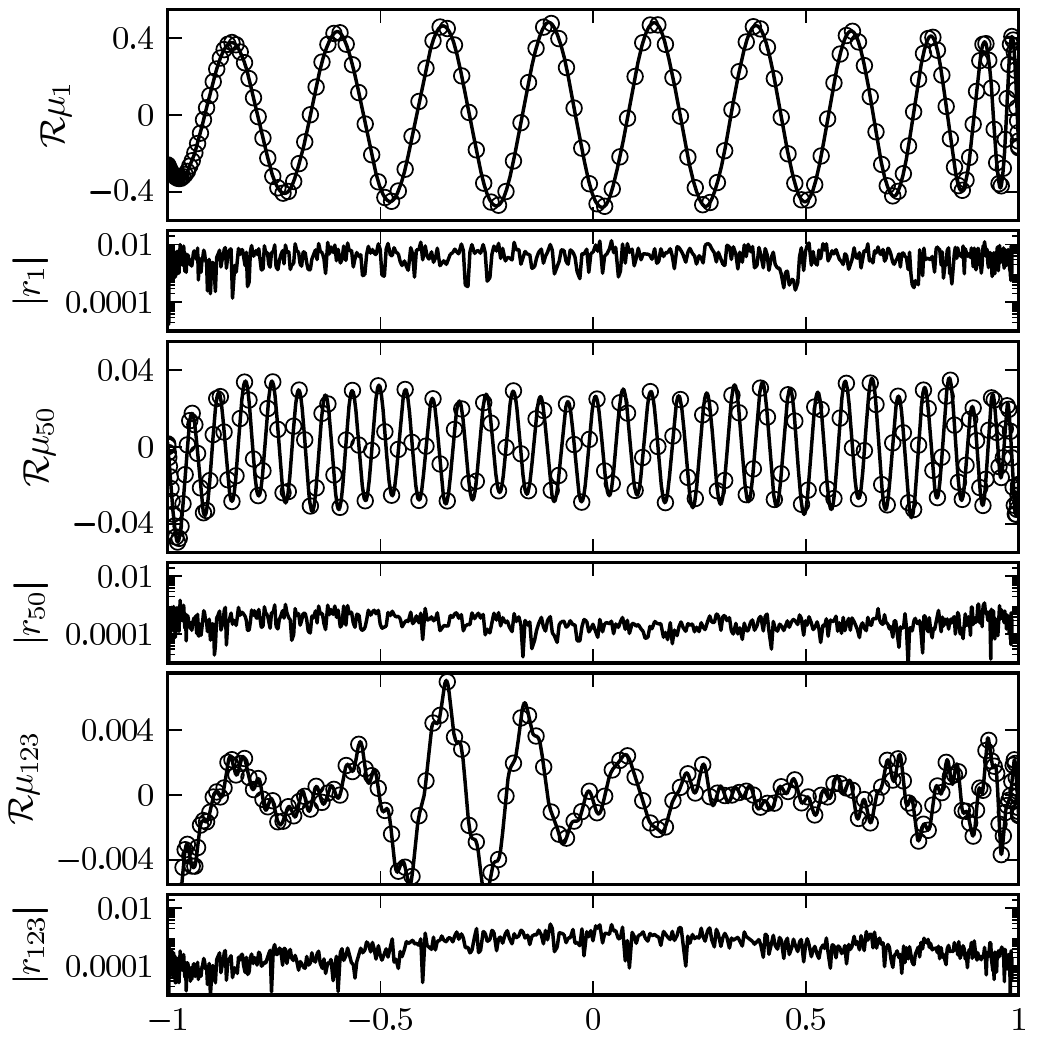}}
\caption{Plotted are the real part of the complex coefficients obtained
from projecting NR waveforms onto the PhenomB SVD basis waveforms ${\bf u}_1$ (top panel),
${\bf u}_{50}$ (third panel), and ${\bf u}_{123}$ (fifth panel). The $x$
axis has been constructed according to \eqref{eq:xspace} and is ideal for
performing a Chebyshev interpolation. Open circles show the
projection coefficients sampled at the collocation points in
\eqref{eq:chebyx} for an $n = 175$ Chebyshev interpolation; solid lines
trace the resultant interpolation for each basis mode. Interpolation is
performed separately on the real and imaginary parts of $\mu_k(x)$ and
combined afterward to obtain the complex function $\mu_k^\prime(x)$. Below
each set of coefficients is the absolute value of the interpolation
residual $|r_k|$ defined in \eqref{eq:resid}.} 
\label{figure:interpcoeff}
\end{figure}

In this space, we find the real and imaginary components, $\mathcal{R}\mu_k(x)$ and
$\mathcal{I}\mu_k(x)$, of the complex projection coefficients to be
oscillatory functions that can roughly be described by a single
frequency. This behaviour is plotted for the basis modes $k = 1$, 50, and
123 in Figure \ref{figure:interpcoeff}. Another trend observed in this
plot is that the projection coefficients become increasingly complex (i.e.\ show less structure) for
higher-order modes. This is a direct result of the increasing complexity of
higher-order basis waveforms themselves. We find that the low-order waveforms are
smoothest while the high-order modes feature many of the
irregularities associated with the multiple frequency components and merger
features of the templates. Though they are more complex, higher-order modes
have smaller singular values and are therefore less important in representing
waveforms. This is evident from the steady decline in
amplitude of the projection coefficients when moving down the different
panels of Figure \ref{figure:interpcoeff}.

We shall use Chebyshev polynomials to interpolate the projection coefficients.
These are a set of orthogonal functions where the $j^{\rm th}$ Chebyshev
polynomial is defined as \begin{align} T_j(x) &\equiv \cos(j\arccos x), & x
&\in [-1,1].  \label{eq:chebyn} \end{align} The orthogonality of Chebyshev
polynomials can be exploited to perform an $n^{\rm th}$ order Chebyshev
interpolation by sampling $\mu_k(x)$ at the $n+1$ so-called collocation points
given by the Gauss-Lobatto Chebyshev nodes \cite{Brutman:1984}
\begin{equation}
x_i = - {\rm cos}\left( \frac{i \pi}{n} \right),
\label{eq:chebyx} \end{equation}
for $i = 0, 1, \dots, n$. 


In general, the interpolation will not be exact and some residual, $r_k(M)$, will be introduced:
\begin{equation}
r_k(M) \equiv \mu_k^\prime(M) - \mu_k(M).
\label{eq:resid}
\end{equation} 
Here $\mu_k(M)$ is the actual coefficient of ${\bf h}_{\rm NR}(M)$ projected onto the basis waveform 
${\bf u}_k$, 
\begin{equation}
\mu_k(M) \equiv \langle {\bf h}_{\rm NR}(M), {\bf u}_k \rangle,
\end{equation}
and $\mu_k^\prime(M)$ is the coefficient obtained after interpolation.  The
new waveform family is expressed numerically as a function of mass through
the relation
\begin{equation}
\label{eq:hintp}
{\bf h}_{\rm intp}(M) = \sum_{k=1}^{N'} \mu_k^\prime(M) {\bf u}_k =
\sum_{k=1}^{N'} \left[\mu_k(M)+r_k(M)\right]{\bf u}_k,
\end{equation}
where the subscript ``intp" reminds the reader that this is computed from an interpolation over $\mu_k$.
An interpolated waveform of total mass $M$ can be compared to the original NR waveform (which we consider
to be the ``true" signal), where the latter is expressed as
\begin{equation}
{\bf h}_{\rm NR}(M)=\sum_{k=1}^{2N} \mu_k(M){\bf u}_k +{\bf h}_\perp(M),
\end{equation}
with ${\bf h}_\perp(M)$ denoting the component of ${\bf h}_{\rm NR}(M)$ that is
orthogonal to the SVD basis (i.e.\ orthogonal to all PhenomB waveforms
in the template bank).
${\bf h}_{\rm NR}(M)$ differs from ${\bf h}_{\rm intp}(M)$ by an amount
\begin{align}
\delta{\bf h}&\equiv {\bf h}_{\rm intp}(M)-{\bf h}_{\rm NR}(M)\nonumber \\
\label{eq:def-dh}
&= \sum_{k=1}^{N'}r_k(M){\bf u}_k 
\;-\!\! \sum_{k=N'+1}^{2N}\mu_k(M){\bf u}_k-{\bf h}_\perp(M)
\end{align}

To compute the impact of the various approximations influencing \eqref{eq:def-dh}, we calculate
the overlap between the interpolated waveform, and the exact waveform,
${\cal O}\left[{\bf h}_{\rm intp}(M),{\bf h}_{\rm NR}(M)\right]$.
To begin this calculation, it is useful to consider
the square of the overlap,
\begin{equation}\label{eq:OverlapSquared}
{\cal O}({\bf h}+\delta{\bf h},{\bf h})^2 
= \frac{\langle{\bf h},{\bf h}+\delta{\bf h}\rangle
\langle{\bf h}+\delta{\bf h},{\bf h}\rangle}
{\langle{\bf h},{\bf h}\rangle
\langle{\bf h}+\delta{\bf h},{\bf h}+\delta{\bf h}\rangle},
\end{equation}
where we have dropped the explicit mass-dependence and subscripts
for convenience. 
Using $\langle{\bf h},{\bf h}\rangle=1$ and Taylor-expanding
the right-hand-side of \eqref{eq:OverlapSquared}
 to second order in $\delta{\bf h}$, we find
\begin{equation}
{\cal O}({\bf h}+\delta{\bf h},{\bf h})^2 
= 1 - \langle\delta{\bf h},\delta{\bf h}\rangle
    + \langle\delta{\bf h},{\bf h}\rangle\langle{\bf h},\delta{\bf h}\rangle.
\end{equation}
To second order in $\delta{\bf h}$, the mismatch is therefore
\begin{equation}\label{eq:MismatchInterp}
{\cal M}({\bf h}+\delta{\bf h}, {\bf h}) 
= \frac{1}{2} \langle\delta{\bf h},\delta{\bf h}\rangle
- \frac{1}{2} \langle\delta{\bf h},{\bf h}\rangle\langle{\bf h},\delta{\bf h}\rangle.
\end{equation}
We note that the right-hand-side of \eqref{eq:MismatchInterp} can be written as
$\frac{1}{2}\langle\delta{\bf h}_\perp,\delta{\bf h}_\perp\rangle$, where $\delta{\bf h}_\perp$ is the part of $\delta{\bf h}$ orthogonal to ${\bf h}$,
\begin{equation}
\delta{\bf h}_\perp = \delta{\bf h} - \langle\delta{\bf h},{\bf h}\rangle{\bf h}.
\end{equation}
However, for simplicity, we proceed by dropping the last term in
\eqref{eq:MismatchInterp}:
\begin{equation}
{\cal M}({\bf h}+\delta{\bf h}, {\bf h}) 
\le \frac{1}{2} \langle\delta{\bf h},\delta{\bf h}\rangle.
\end{equation}
Using \eqref{eq:def-dh}, this gives
\begin{multline}
\label{eq:InterpolationMismatch1}
{\cal M}\left[{\bf h}_{\rm intp}(M), {\bf h}_{\rm NR}(M)\right] \le \\
\frac{1}{2}\sum_{k=1}^{N'} \left|r_k(M)\right|^2 +
\frac{1}{2}\sum_{k=N'+1}^{2N}\left|\mu_k(M)\right|^2 + \frac{1}{2}|{\bf
h}_\perp|^2 
\end{multline}
We thus see three contributions to the total mismatch: (i) the interpolation
error, $\sum_{k=1}^{N'}|r_k(M)|^2$; (ii) the truncation error from the
discarded waveforms of the reduced basis, $\sum_{k=N'+1}^{2N}|\mu_k(M)|^2$;
(iii) the failure of the SVD basis to represent the NR waveform, $|{\bf
h}_\perp|$.  The sum of the last two terms, which together make up the
representation error, is traced by the dashed line in Figure \ref{figure:rec}.
The goal for our new waveform family is to have an interpolation error that is
negligible compared to the representation error.

\begin{figure}
\resizebox{\linewidth}{!}{\includegraphics{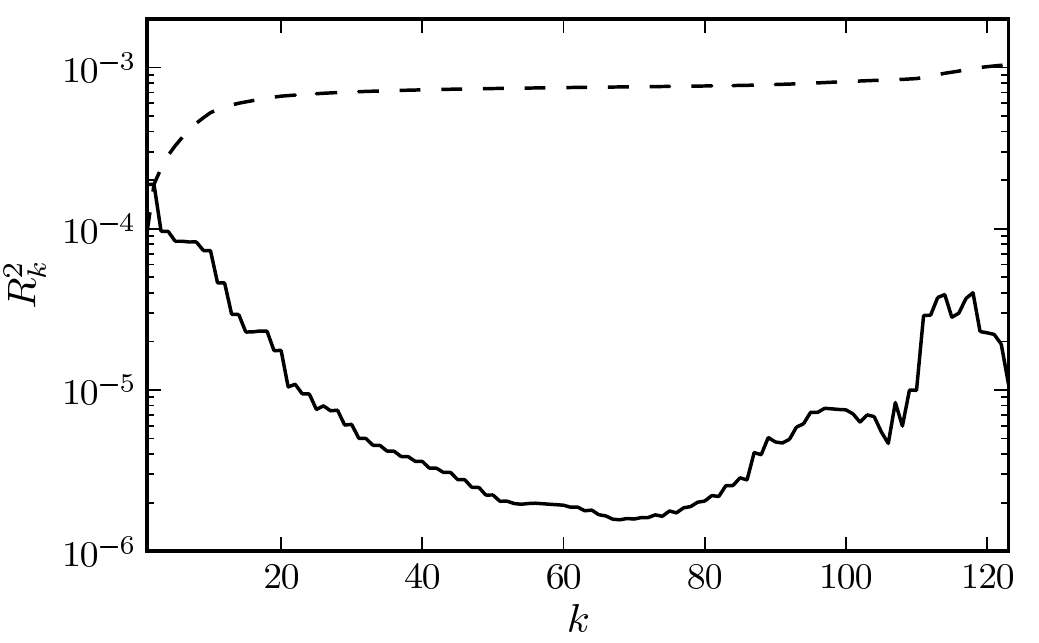}}
\caption{The solid line traces the interpolation error $R_k^2$ in \eqref{eq:rkmax} maximized
over mass for each mode $k$. This is used in \eqref{eq:InterpolationMismatch2} to define
an upper bound to the interpolation error arising in \eqref{eq:InterpolationMismatch1}. The 
cumulative sum in the latter expression is traced by the dashed curve and shows that the 
interpolation error is dominated by the lowest-order basis waveforms.}
\label{figure:rk}
\end{figure}

To remove the mass-dependence of interpolation error in \eqref{eq:InterpolationMismatch1}, 
we introduce the maximum interpolation error of each mode,
\begin{equation}
R_k\equiv \max_M |r_k(M)|.
\label{eq:rkmax}
\end{equation}
This allows the bound
\begin{equation}
\label{eq:InterpolationMismatch2}
\frac{1}{2} \sum_{k=1}^{N'} R_k^2 \geq \frac{1}{2} \sum_{k=1}^{N^\prime}|r_k(M)|^2
\end{equation}
to place an upper limit on the error introduced by interpolation.
Figure \ref{figure:rk} plots $R_k^2$ as a function of mode-number $k$ as well as the 
cumulative sum $\sum_{k=1}^{N^\prime} R_k^2/2$. The data pertains to an interpolation
performed using $n = 175$ Chebyshev polynomials on the reduced SVD basis containing
the frist $N^\prime = 123$ of $2N = 254$ waveforms. In this case, we find the interpolation
error to be largely dominated by the lowest-order modes and also partially by the
highest-order modes. Interpolated coefficients for various modes are plotted in Figure
\ref{figure:interpcoeff} and help to explain the features seen in Figure \ref{figure:rk}. 
In the first place, interpolation becomes increasingly more difficult for higher-order modes due 
to their increasing complexity. This problem is mitigated by the fact that high-order modes are 
less important for representing waveforms, as evidenced by the diminishing amplitude of
projection coefficients. Although low-order modes are much smoother and thus easier to
interpolate, their amplitudes are considerably larger meaning that interpolation errors are
amplified with respect to high-order modes. 


\begin{figure}
\resizebox{\linewidth}{!}{\includegraphics{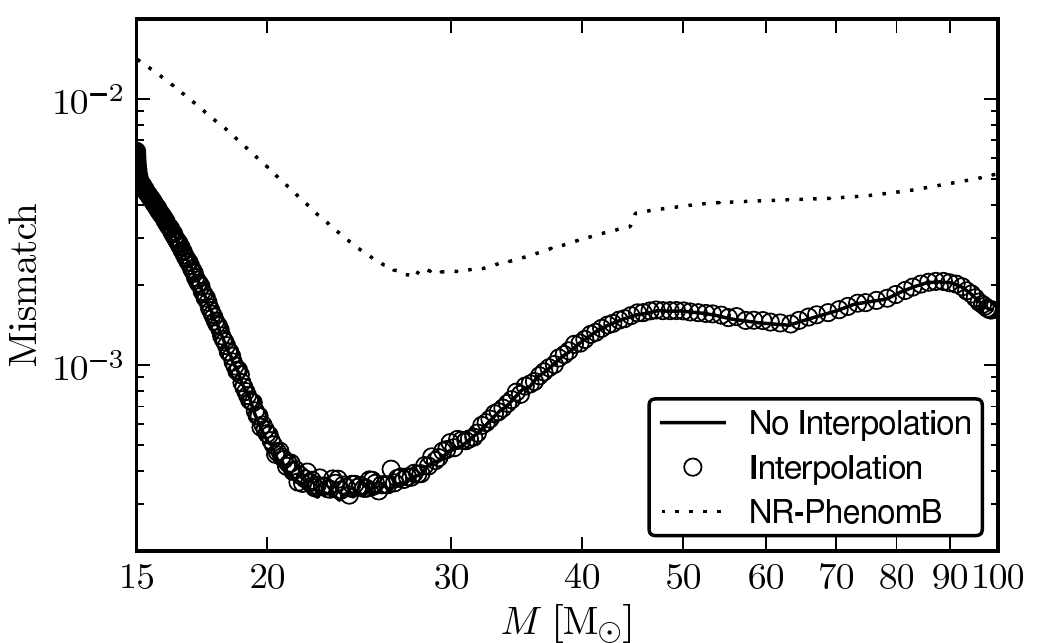}}
\caption{Open circles show the total mismatch of our new waveform family obtained by interpolating
the coefficients of NR waveforms projected onto a reduced SVD basis of PhenomB waveforms.
To highlight the error introduced by interpolation, the solid curve traces only the mismatch of NR waveforms
represented by the reduced basis (i.e.\ the dashed curve in Figure \ref{figure:rec}). The total interpolation
mismatch is lower than the dotted line tracing the mass-optimized mismatch between NR and PhenomB
waveforms (i.e.\ the solid line in the top panel of Figure \ref{figure:bias}) and demonstrates the ability
of SVD to boost the accuracy of the PhenomB family.}
\label{figure:interperror}
\end{figure}

\eqref{eq:InterpolationMismatch1} summarizes the three components adding to
the final mismatch of our interpolated waveform family. Their total contribution
can be computed directly from the interpolated coefficients in a manner
similar to \eqref{eq:recerror}:
\begin{equation}
\mathcal{M}[{\bf h}_{\rm intp}(M), {\bf h}_{\rm NR}(M)] =
1 - \sqrt{\sum_{k=1}^{N^\prime} \mu_k^\prime(M) \mu_k^{\prime^*}(M)}.
\label{eq:InterpolationMismatch3}
\end{equation}
In the case of perfect interpolation for which $\mu^\prime_k(M) = \mu_k(M)$,
\eqref{eq:hintp} and \eqref{eq:InterpolationMismatch3} reduce to
\eqref{eq:hproj} and \eqref{eq:recerror} respectively, and the total mismatch is simply
the representation error of the reduced basis. 

In Figure \ref{figure:interperror} open circles show the total mismatch
\eqref{eq:InterpolationMismatch3} between our interpolated waveform family and
the true NR waveforms for various masses. Also plotted is the NR representation
error without interpolation and the mismatch between NR and PhenomB waveforms
minimized over mass.  We see that interpolation introduces only small
additional mismatch to the interpolated waveform family, and remains well below
the optimized NR-PhenomB mismatch. This demonstrates the efficacy of using SVD
coupled to NR waveforms to generate a \emph{faithful} waveform family with
improved accuracy over the \emph{effectual} PhenomB family that was originally
used to create templates. This represents a general scheme for improving
phenomenological models and presents an interesting new opportunity to enhance
the matched-filtering process employed by LIGO. 

\section{Higher Dimensions}
\label{sec:2d}

\begin{figure*}
\resizebox{\linewidth}{!}{\includegraphics{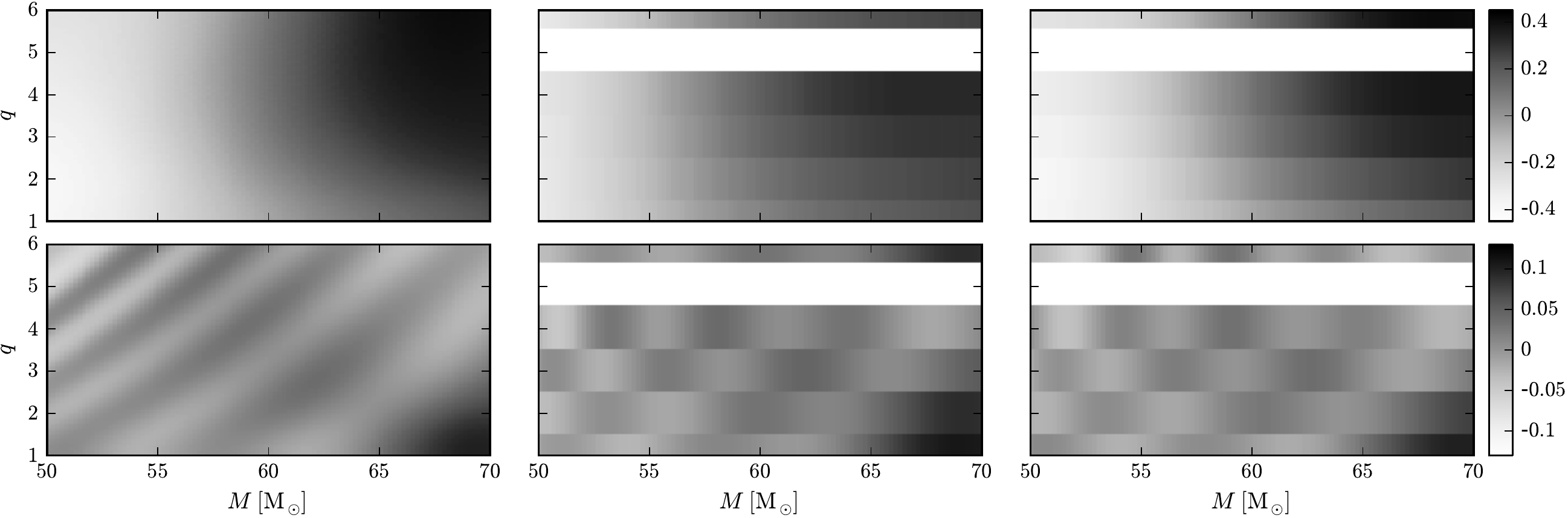}}
\caption{(left panels) Real components of the smoothed PhenomB projection coefficients
$\tilde{\mu}_3(M,q)$ (top) and $\tilde{\mu}_{16}(M,q)$ (bottom) for
mass-ratios $1 \leq q \leq 6$ and total masses $\unit{50}{\Msun} \leq M
\leq \unit{70}{\Msun}$. (middle panels) Real components of the smoothed NR projection
coefficients $\tilde{\mu}_3(M,q)$ (top) and $\tilde{\mu}_{16}(M,q)$
(bottom) for mass-ratios  $q$ = \{1, 2, 3, 4, 6\} and total mass
$\unit{50}{\Msun} \leq M \leq \unit{70}{\Msun}$. (right panels) Real components of 
the smoothed PhenomB projection coefficients $\tilde{\mu}_3(M,q)$ (top) and $\tilde{\mu}_{16}(M,q)$ (bottom) 
coarsened to the same set of mass-ratios as the middle panels. 
For both the middle and rightmost panels an artificial row of white space has been plotted for $q=5$ in order to 
ease comparison with the leftmost panels.} 
\label{figure:coeff2d}
\end{figure*}

So far, we have focused on the total mass axis of parameter space. As already discussed, 
this served as a convenient model-problem, because the $q=1$ NR waveform 
can be rescaled to any total mass, so that we are able to compare against the ``correct'' 
answer. The natural extension of this work is to expand into higher dimensions
where NR waveforms are available only at certain, discrete mass-ratios $q$. In this
section we consider expanding our approach of interpolating NR projection 
coefficients from a two-dimensional template bank containing unequal-mass waveforms.

We compute a template bank of PhenomB waveforms covering mass-ratios 
$q$ from 1 to 6 and total masses $\unit{50}{\Msun} \leq M \leq  \unit{70}{\Msun}$.
This mass range is chosen to facilitate comparison with previous work done by
\cite{cannon/etal:2011c}. For the two-dimensional case the construction of a template bank
is no longer as straightforward as before due to the additional degree of freedom associated
with varying $q$. One method that
has been advanced for this purpose is to place templates hexagonally on the
waveform manifold \cite{cokelaer:2007}. Using this procedure we find $N =
16$ templates are required to satisfy a minimal match of 0.97.

Following the waveform preparation of \cite{cannon/etal:2011c}, templates
are placed in the rows of a matrix ${\bf G}$ with real and imaginary
components filled in alternating fashion where the whitened waveforms are
arranged in such a way that their peak amplitudes are aligned. The
waveforms are sampled for a total duration of $\unit{2}{\second}$ with
uniform spacing $\Delta t = \unit{2^{-15}}{\second}$ so that
$\unit{16}{\mebi\byte}$ of memory is required to store the contents of
${\bf G}$ if double precision is desired.  Application of \eqref{eq:SVD}
transforms the 16 complex-valued waveforms into 32 real-valued orthonormal
basis waveforms.

The aim is to sample the coefficients of NR waveforms projected onto 
the SVD basis of PhenomB waveforms using mass-ratios for which NR data exists, 
and then interpolate amongst these to construct a numerical waveform family that
can be evaluated for arbitrary parameters. This provides a method for evaluating
full IMR waveforms for mass-ratios that have presently not been simulated. To
summarize, we take some NR waveform, ${\bf h}_{\rm NR}(M,q)$, or total mass $M$
and mass-ratio $q$, and project it onto the basis waveform ${\bf u}_k$ in order to 
obtain
\begin{equation}
\mu_k(M, q) = \langle {\bf h}_{\rm NR}(M, q), {\bf u}_k \rangle.
\label{eq:muk2d}
\end{equation}
Next we apply some two-dimensional interpolation scheme on \eqref{eq:muk2d}
to construct continuous functions $\mu_k^\prime(M,q)$ that can be evaluated for 
arbitrary values of $M$ and $q$ bounded by the regions of the template bank.
The interpolated waveform family is given numerically by the form:
\begin{equation}
{\bf h}_{\rm intp}(M, q) = \sum_{k=1}^{N^\prime} \mu_k^\prime(M, q) {\bf u}_k.
\label{eq:hint2d}
\end{equation}

As before, the interpolation process works best if we can develop a scheme for
which the projection coefficients are smoothly varying functions of $M$ and
$q$. Following the procedure described in \cite{cannon/etal:2011c}, the complex
phase of the first mode is subtracted from all modes:
\begin{equation}
\tilde{\mu}_k(M,q) \equiv \ee^{-\aye \arg[\mu_1(M, q)]}\mu_k(M,q).
\label{eq:musmooth}
\end{equation}
To motivate why \eqref{eq:musmooth} might be useful, let us consider modifying
the PhenomB waveform family with a parameter-dependent complex phase
$\Phi(M,q)$: \begin{equation} {\bf h}_{\rm PB}(M,q) \to  \ee^{\aye\Phi(M,q)}
{\bf h}_{\rm PB}(M,q).  \end{equation} When constructing a template bank, or
when using a template bank, such a complex phase $\Phi(M,q)$ is irrelevant,
because the waveforms are always optimized over a phase-shift.  However,
$\Phi(M,q)$ will appear in the projection coefficients, \eqref{eq:muk2d},
\begin{equation}
\mu_k(M,q) \to \ee^{\aye\Phi(M,q)}\mu_k(M,q).
\end{equation}
Therefore, if one had chosen a function $\Phi(M,q)$ with fine-scale structure,
this structure would be inherited by the projection coefficients $\mu_k(M,q)$.
For traditional uses of waveform families the overall complex phase $\Phi(M,q)$
is irrelevant, and therefore, little attention may have been paid to how it
varies with parameters $(M,q)$.  The transformation \eqref{eq:musmooth} removes
the ambiguity inherent in $\Phi(M,q)$ by choosing it such that
$\arg\tilde\mu_1(M,q)=0$.  This choice ties the complex phase to the physical
variations of the $\mu_1$ coefficient, and does therefore eliminate all
unphysical phase-variations on finer scales. 

In the leftmost panels of Figure \ref{figure:coeff2d} we plot the real part of
the smoothed coefficients $\tilde{\mu}_k(M,q)$ for PhenomB waveforms projected
onto the basis modes $k = 3$ and $k = 16$. The middle panels show the same
thing except using the NR waveforms evaluated at the set of mass-ratios $q$ =
\{1, 2, 3, 4, 6\} for which we have simulated waveforms.  Obviously, the
refinement along the $q$ axis is much finer for the PhenomB waveforms since
they can be evaluated for arbitrary mass-ratio, whereas we are limited to
sampling at only 5 discrete mass-ratios for NR waveforms. For comparison
purposes, the rightmost panels of Figure \ref{figure:coeff2d} show the PhenomB
projection coefficients coarsened to the same set of mass-ratios for which the
NR waveforms are restricted to.

We find the same general behaviour as before that low-order modes display the smoothest
structure, while high-order modes exhibit increasing complexity. A plausible interpolation scheme
would be to sample $\tilde{\mu}_k$ for NR waveforms of varying mass for constant mass ratio
(i.e.\ as we have done previously) and then stitch these together across the $q$ axis. Since the
projection coefficients in Figure \ref{figure:coeff2d} show sinusoidal structure
they must be sampled with at least the Nyquist frequency along both axes. However, 
looking at the middle and rightmost panels it appears as though this is not yet possible given the present
set of limited NR waveforms. At best the 5 available mass-ratios are just able to sample at the Nyquist 
frequency along the $q$ axis for high-order modes. In order to achieve a reasonable interpolation from
these projection coefficients the current NR data thus needs to be appended with more mass-ratios. 
Based on the left panels of Figure \ref{figure:coeff2d} a suitable choice would be to double the 
current number of mass-ratios to include $q$ = \{1.5, 2.5, 3.5, 4.5, 5, 5.5\}.
Hence, though it is not yet practical to generate an interpolated waveform family 
using the SVD boosting scheme applied to NR waveforms, the possibility remains open as more NR waveforms
are generated. 

\section{Discussion}
\label{sec:discussion}

We have shown that SVD can be used to improve the representation of
NR waveforms from a PhenomB template bank. A reasonably reduced SVD
basis was able to reduce mismatch by a factor of five compared to PhenomB
waveforms optimized over mass. There was also no mass-bias associated with 
the SVD basis and therefore no optimization over physical parameters required. 
This occurs because SVD unifies a range of waveform
structure over an extended region of parameter space so that any biases become 
blended into its basis. SVD therefore represents a generalized scheme through which
phenomenological waveform families can be de-biased and enhanced for use as matched-filter
templates.  

We were able to calibrate an SVD basis of PhenomB templates against NR waveforms in
order to construct a new waveform family with improved accuracy. This was accomplished by interpolating the
coefficients of NR waveforms projected onto the PhenomB basis. Only marginal error
was introduced by the interpolation scheme and the new waveform family provided a more
faithful representation of the ``true'' NR signal compared to the original PhenomB model.
This was shown explicitly for the case of equal-mass, zero-spin binaries. We proceeded to
investigate the possibility of extending this approach to PhenomB template banks containing
unequal-mass waveforms. At present, however, this method is not yet feasible since the current
number of mass-ratios covered by NR simulations are unable to sample the projection coefficients
with the Nyquist frequency. This method will improve as more NR waveforms are simulated and should
be sufficient if the current sampling rate of mass-ratios were to double.

\acknowledgments We thank Ilana MacDonald for preparing the hybrid waveforms
used in this study. KC, JDE and HPP gratefully acknowledge the support of the
National Science and Engineering Research Council of Canada, the Canada
Research Chairs Program, the Canadian Institute for Advanced Research, and
Industry Canada and the Province of Ontario through the Ministry of Economic
Development and Innovation. DK gratefully acknowledges the support of the Max
Planck Society.

\end{document}